\documentclass[preprint,showpacs,preprintnumbers,amsmath,amssymb,aps,nofootinbib,floatfix]{revtex4}

\usepackage{graphicx}% Include figure files
\usepackage{dcolumn}% Align table columns on decimal point
\usepackage{bm}% bold math
\usepackage{graphicx}
\usepackage[figuresright]{rotating}
\usepackage{dcolumn}% Align table columns on decimal point
\usepackage{bm}% bold math

\def\chpt{\raise0.4ex\hbox{$\chi$}PT}
\def\schpt{S\raise0.4ex\hbox{$\chi$}PT}
\def\figref#1{Fig.~\ref{fig:#1}}

\def\secref#1{Sec.~\ref{sec:#1}}

\def\leftvec{\raise1.5ex\hbox{$\leftarrow$}\kern-.85em}
\def\half{{\scriptstyle \raise.2ex\hbox{${1\over2}$}}}
\def\threehalves{{\scriptstyle \raise.15ex\hbox{${3\over2}$}}}
\def\third{{\scriptstyle \raise.15ex\hbox{${1\over3}$}}}
\def\third{{\scriptstyle \raise.15ex\hbox{${1\over3}$}}}
\def\twothirds{{\scriptstyle \raise.15ex\hbox{${2\over3}$}}}
\def\fourth{{\scriptstyle \raise.15ex\hbox{${1\over4}$}}}
\def\gtwid{\raise.3ex\hbox{$>$\kern-.75em\lower1ex\hbox{$\sim$}}}
\def\ltwid{\raise.3ex\hbox{$<$\kern-.75em\lower1ex\hbox{$\sim$}}}

\def\cO{{\cal O}}

\def\eqn#1{\label{eq:#1}}

\def\eq#1{Eq.~(\ref{eq:#1})}
\def\eqs#1#2{Eqs.~(\ref{eq:#1}) and (\ref{eq:#2})}

\def\prd#1{Phys.\ Rev.\ D {\bf #1}}

\def\fermilabtwo#1{Nucl.\ Phys.\ {\bf B} (Proc.\ Suppl.) {\bf 140}, #1 (2005)}

\def\MeV{{\rm Me\!V}}

\def\ie{{\it i.e.},\ }
\def\eg{{\it e.g.},\ }
\def\et{{\it et al.}}

\begin{document}

% declarations for front matter
\title{Order of the Chiral and Continuum Limits in Staggered Chiral Perturbation 
Theory}
\author{C.~Bernard} 
\affiliation{Department of Physics, Washington University, St.~Louis, MO 63130, USA}

\date{\today}

\begin{abstract}
D\"urr and Hoelbling recently observed that the 
continuum and chiral limits do not commute in the two dimensional, one flavor, Schwinger model with
staggered fermions. I point out that such lack of commutativity can also
be seen in four-dimensional staggered chiral perturbation theory (\schpt) in quenched
or partially quenched quantities constructed to be particularly sensitive to the chiral limit.
Although the physics involved in the \schpt\ examples is quite different 
from that in the Schwinger model, neither
singularity seems to be connected to the trick of taking the
n${}^{\rm th}$ root of the fermion determinant to remove unwanted degrees of
freedom (``tastes''). 
Further, I argue that the singularities in \schpt\ are absent
in most commonly-computed quantities in the unquenched (full) QCD case 
and do not imply any unexpected
systematic errors in recent MILC calculations with staggered fermions.
\end{abstract}

\maketitle

\section{Introduction}
\label{sec:intro}
The one-flavor Schwinger model in two dimensions (2D) has recently been studied by
D\"urr and Hoelbling \cite{Durr:2004ta}
using both overlap and staggered fermions.  The square root of the
staggered determinant is used 
to eliminate the extra taste degree of freedom. This is 
basically the same trick that is employed
in current MILC three-flavor QCD simulations \cite{FPI04,strange-mass,MILC_SPECTRUM2,BIG_PRL}
(the fourth root is required in 4D).  Although many of the results in
Ref.~\cite{Durr:2004ta} are encouraging for the use of ``rooted'' staggered quarks,
there is one disturbing feature:
The chiral and continuum limits of $\langle\bar\psi\psi\rangle$ do not commute.  If the continuum limit is
taken first, the known continuum result is reproduced.
But if the chiral limit is first taken for the
staggered quarks, the continuum limit then disagrees with the exact result. 
This leads to three key questions: 
\begin{itemize}
\item[$\bullet$]{} Do similar singularities appear in 4D QCD with staggered fermions?
\item[$\bullet$]{} If they do appear, do such singularities induce uncontrolled systematic errors in
the results reported in, \eg  Refs.~\cite{FPI04,strange-mass,MILC_SPECTRUM2,BIG_PRL}?
\item[$\bullet$]{} Are the singularities the result of the rooting procedure?
\end{itemize}

Here, I address these questions for 4D QCD in the context of 
staggered chiral perturbation theory (\schpt) \cite{LEE-SHARPE,SCHPT1,SCHPT2,Billeter:2004wx,Sharpe:2004is,Aubin:2004xd}.
In \schpt, singular quantities for which the chiral and continuum limits fail to commute can
easily be found in the quenched or partially quenched cases. 
As explained in \secref{examples},
the way the noncommutativity comes about is simple:
Taste violations split the masses of mesons, and the splitting
becomes the dominant effect in the meson masses as the chiral limit is approached.
When the physics is particularly sensitive to the chiral regime, 
the chiral and continuum limits do not commute.  
Quenching or partial quenching leads to such sensitivity through 
double poles in neutral meson propagators, which 
enhance the infrared (IR) regime in loop diagrams. 
In the context of \schpt, the lack of commutativity has nothing directly to do with the issue of taking
the root of the staggered determinant, and occurs in normal, ``unrooted,'' staggered
theories also. 

In \secref{IR}, I discuss the issue of commutativity of limits for unquenched
\schpt\ (``full \schpt'').  Defining safe quantities as
those for which the chiral and continuum limits commute, I argue that
most standard physical quantities  are safe because
the full theory has no double poles and is therefore better behaved in the IR.  In this sense it is 
very similar to full continuum chiral perturbation theory (\chpt).
The safe quantities include those commonly determined in simulations,
such as $f_\pi$ or the ratio of squared Goldstone meson mass to quark mass.
However, the shorthand statement ``standard quantities are safe'' is misleading, since
is in fact simple to write down other quantities
for which the limits do not commute in full \schpt.  A rough guide is that quantities that
are finite in the chiral limit of full, continuum \chpt\ are likely to be safe in this
sense, although this guide itself has exceptions. Furthermore, 
there is at this point no proof, but only an
intuitive argument for why the singularities are absent to all orders for the
standard quantities in the full case.  
However, as discussed in \secref{errors}, even if one assumes that the argument breaks down and
singularities actually appear at some higher 
order, the associated new errors in previously computed quantities would be negligible.
 
Finally, in \secref{remarks}, I mention two other examples of lack of commutativity, one 
with Wilson fermions and one in the quenched 2D Schwinger model. I point out that the known examples seem to indicate
that the rooting procedure and the commutativity issues are independent. I conclude
with a discussion of whether noncommutativity might be seen in topological and $\eta'$
physics in 4D.

\section{Examples from \schpt}
\label{sec:examples}
The simplest example of noncommutativity of limits in \schpt\ occurs in the quenched pion mass, 
where the ``pion'' is made of two degenerate valence
quarks of mass $m_{\scriptscriptstyle V}$.  In continuum quenched \chpt, we have at next-to-leading order (NLO) \cite{QCHPT}:
\begin{equation}\eqn{CQmass}
\left(\frac{M_\pi^2}{2\mu m_{\scriptscriptstyle V}}\right)_{\rm cont} = 1 +\frac{1}{16\pi^2f^2}\; \frac{-2m_0^2}{3} \;
\ln\left(\chi_{\scriptscriptstyle VV}/\Lambda\right) +\cdots  \ ,
\end{equation}
where $\Lambda$ is the chiral scale;
$m_0^2$, the contribution to the $\eta'$ mass from the anomaly;
$f$, the decay constant normalized so that $f\approx 131\MeV$; and 
$\chi_{\scriptscriptstyle VV}$, the tree-level pion mass squared, $\chi_{\scriptscriptstyle VV} = 2\mu m_{\scriptscriptstyle V}$. ($\mu$
is a constant with dimensions of mass.)  The 
$\cdots$ represents less singular terms that do not lead to noncommutativity of limits.
Here, and throughout this paper, I work in the infinite volume limit for simplicity.

The corresponding expression for the Goldstone pion in \schpt\ is \cite{SCHPT1}
\begin{equation}\eqn{SQmass}
\left(\frac{M_\pi^2}{2\mu m_{\scriptscriptstyle V}}\right)_{\rm stag} = 1 +\frac{1}{16\pi^2f^2}\; \frac{-2m_0^2}{3} \;
\ln\left(\chi^I_{\scriptscriptstyle VV}/\Lambda\right) +\cdots  \ ,
\end{equation}
where $\chi^I_{{\scriptscriptstyle VV}}$ is the squared tree-level mass of the taste-singlet pion,
\begin{equation}\eqn{ChiVVI}
\chi^I_{{\scriptscriptstyle VV}} = \chi_{\scriptscriptstyle VV} + a^2 \Delta_I = 2\mu m_{\scriptscriptstyle V} +a^2 \Delta_I \,
\end{equation}
with $a^2\Delta_I$ the splitting of the taste-singlet pion from the
Goldstone pion. In \eq{SQmass} the neglected terms include effects of the taste-violating
hairpins \cite{SCHPT1}.  These are suppressed by $a^2$ and cause no problems with the limits. 

Taking the ratio of lattice to continuum results, we have, to this order,
\begin{equation}\eqn{Qratio}
\frac{\left(M_\pi^2/m_{\scriptscriptstyle V}\right)_{\rm stag} }
{\left(M_\pi^2/m_{\scriptscriptstyle V}\right)_{\rm cont} }
= 1 +\frac{1}{16\pi^2f^2}\;\frac{-2m_0^2}{3} \;
\ln\left(1+\frac{a^2\Delta_I}{\chi_{\scriptscriptstyle VV}}\right) +\cdots \ .
\end{equation}
This ratio clearly goes to 1 if we take the continuum limit, $a\to0$, first.  But it blows
up if the chiral limit ($m_{\scriptscriptstyle V}\to0$, $\chi_{\scriptscriptstyle VV}\to0$) is taken first, 
because the staggered theory
gives the wrong answer in this case.  Of course \chpt\ breaks down once
the correction term gets large.  In this case the breakdown occurs because of the
``quenched chiral log'' in the continuum. The logarithm is cut off in the staggered theory
by the taste splitting, $a^2\Delta_I$.

This quenched example makes my basic point, but it is not very similar 
to the effect seen in Ref.~\cite{Durr:2004ta}.  There, the staggered theory is found
to give a finite result for either order of the limits $m_{\scriptscriptstyle V}\to0$ and $a\to0$;
it is just that the result is incorrect (different from the continuum)
when  $m_{\scriptscriptstyle V}\to0$ is taken first.
An example that has this kind of behavior can be found in the partially quenched 4D theory.\footnote{But
note that the example in Ref.~\protect{\cite{Durr:2004ta}} occurs in the {\em full} 2D theory.}

For simplicity,  I take $N_{\scriptscriptstyle S}$ degenerate sea quarks of mass $m_{\scriptscriptstyle S}$
in the partially quenched
theory, and again consider a pion made from degenerate valence quarks of mass
$m_{\scriptscriptstyle V}$.  In the continuum, the pion mass at NLO is \cite{Sharpe:1997by}
\begin{equation}\eqn{CPQmass}
\left(\frac{M_\pi^2}{2\mu m_{\scriptscriptstyle V}}\right)_{\rm cont} = 1 +\frac{1}{16\pi^2f^2}\; \frac{2}{N_{\scriptscriptstyle S}} \;
\left(2\chi_{\scriptscriptstyle VV} -\chi_{\scriptscriptstyle SS}\right)\ln\left(\chi_{\scriptscriptstyle VV}/\Lambda\right) +\cdots \ ,
\end{equation}
where $\chi_{\scriptscriptstyle SS}=2\mu m_{\scriptscriptstyle S}$, and $\cdots$ represents analytic terms.

In the staggered case, consider
$N_{\scriptscriptstyle F}$ degenerate dynamical staggered fields with mass $m_{\scriptscriptstyle S}$.
I define $N_{\scriptscriptstyle S}$ here as the number of sea quarks in 
the continuum limit: $N_{\scriptscriptstyle S}=N_{\scriptscriptstyle F}$ 
if the fourth root of the determinant is taken; while $N_{\scriptscriptstyle S} =4N_{\scriptscriptstyle F} $ if it is not.
The result has the same form when written in terms of $N_{\scriptscriptstyle S}$, whether or not
the root is taken.  Putting sea and valence  quarks separately degenerate in 
Eq.~(48) of Ref.~\cite{SCHPT1}, one easily arrives at the one-loop mass of the
Goldstone pion:
\begin{equation}\eqn{SPQmass}
\left(\frac{M_\pi^2}{2\mu m_{\scriptscriptstyle V}}\right)_{\rm stag} = 1 +\frac{1}{16\pi^2f^2}\; \frac{2}{N_{\scriptscriptstyle S}} \;
\left(2\chi^I_{\scriptscriptstyle VV} -\chi^I_{\scriptscriptstyle SS}\right)\ln\left(\chi^I_{\scriptscriptstyle VV}/\Lambda\right) +\cdots \ ,
\end{equation}
where $\chi^I_{\scriptscriptstyle SS}$ is defined analogously to $\chi^I_{\scriptscriptstyle VV}$, \eq{ChiVVI}, and $\cdots$ represents analytic terms and the effects of taste-violating hairpins, which again
cause no problem with the limits.

If the continuum limit is taken first, \eq{SPQmass} reproduces \eq{CPQmass}, so the two will
give identical results no matter how the chiral limit is subsequently taken.  On the other
hand, if first we take the valence chiral limit ($m_{\scriptscriptstyle V}\to0$, with $m_{\scriptscriptstyle S}$ fixed),
the lack of commutativity of limits exactly parallels the quenched case,
with $\chi_{\scriptscriptstyle SS}$ playing the role of $m_0^2$.  
As before, $\cO(a^2)$ terms in the staggered theory cut off a chiral log that is divergent in the
continuum theory. 

For my purposes, a more interesting chiral limit occurs when 
$m_{\scriptscriptstyle V}$ and $m_{\scriptscriptstyle S}$ both approach 0, but with $m_{\scriptscriptstyle S}$ vanishing much more slowly, so that
\begin{equation}\eqn{VSchirallimit}
\chi_{\scriptscriptstyle SS} \sim \frac{-C}{\ln(\chi_{\scriptscriptstyle VV}/\Lambda)} 
\end{equation}
as $\chi_{\scriptscriptstyle VV}\to 0$, with $C$ a positive constant.  In this limit (with $a$ fixed in the staggered case)
\begin{eqnarray}\eqn{CPQmasslimit}
\left(\frac{M_\pi^2}{2\mu m_{\scriptscriptstyle V}}\right)_{\rm cont} &\to & 1 +\frac{1}{16\pi^2f^2}\, \frac{2}{N_{\scriptscriptstyle S}}\, C \\
\left(\frac{M_\pi^2}{2\mu m_{\scriptscriptstyle V}}\right)_{\rm stag} &\to &  1 +\frac{1}{16\pi^2f^2}\, \frac{2}{N_{\scriptscriptstyle S}} \,
a^2\Delta_I\ln\left(a^2\Delta_I/\Lambda\right) +\cdots \ ,
\end{eqnarray}
where $\cdots$ represents additional $\cO(a^2\ln(a^2))$ or $\cO(a^2)$ terms, coming from taste-violating hairpins
or taste-violating analytic terms. 
If we now take the continuum limit,
\begin{eqnarray}\eqn{SPQmasslimit}
\left(\frac{M_\pi^2}{2\mu m_{\scriptscriptstyle V}}\right)_{\rm stag} \to   1  \ ,
\end{eqnarray}
in disagreement with \eq{CPQmasslimit}. On the other hand, the staggered theory clearly
reproduces \eq{CPQmasslimit} if the continuum limit is taken first in \eq{SPQmass}.

Note that the noncommutativity of limits has nothing to do with the fourth-root prescription,
and has exactly the same form with or without the fourth root.  Of course, for fixed
number $N_{\scriptscriptstyle F}$ of dynamical staggered fields, the effective number of
continuum sea quarks $N_{\scriptscriptstyle S}$ depends on whether the
root is taken, so we could say in that sense that the rooting prescription trivially affects the
the strength of the singularity, but not its existence.

\section{Commutativity in full \schpt}
\label{sec:IR}
The lack of commutativity of limits appears in the computed one-loop \schpt\ 
quantities \cite{SCHPT1,SCHPT2,Aubin:2004xd} only in the quenched or partially
quenched cases, which are enhanced in the IR relative to the full case
and therefore especially sensitive to the chiral
limit. This is due to the double pole 
structure that shows up in flavor-neutral meson propagators.
In the full QCD case, there are only single poles, and I therefore do not
expect noncommutativity to be generic in this case. For example,
if we first go to full QCD ($m_{\scriptscriptstyle V}=m_{\scriptscriptstyle S}\equiv m$; 
$\chi_{\scriptscriptstyle VV}= \chi_{\scriptscriptstyle SS}\equiv \chi$) in \eqs{CPQmass}{SPQmass},
the NLO corrections then always vanish in the subsequent $m\to0$,
$a\to 0$ limits,
independent of their order.  

The absence of the commutativity problem in most full QCD quantities seems to be a general
feature of \schpt. It is equivalent to the statement that the limit $m\to0$ at fixed $a$
is smooth in full \schpt, \ie that there are no IR divergences on shell in this limit.   
I believe this to be the case because,  taking the additional limit $a\to0$, one recovers the
massless continuum \chpt.  Ordinary chiral power counting in this limit
shows that IR divergences are absent:  The derivative couplings suppress IR contributions
and make loop effects finite even in the presence of massless particles.\footnote{Note that
one needs to go on shell so that derivatives acting on external lines also suppress the
diagrams in the massless limit.} If we now turn on the lattice-dependent ($\cO(a^2)$ and higher) vertices
of \schpt, these act as effective mass terms and should not induce IR divergences that were
previously absent. Power counting applied to \schpt\ then implies that any logarithms of $a^2$
that appear in the $m\to0$ limit are ``protected'' by powers of $a^2$, so they cause
no problem in the subsequent $a\to0$ limit.

Full \schpt\ in the chiral limit is in fact closely analogous to 
full continuum \chpt\ with massless up and down quarks but a nonvanishing
strange quark mass. In both cases there are
terms giving mass to some, but not all, of the pseudoscalar mesons: The pions remain massless
in the continuum example; while the taste-$\xi_5$ (Goldstone) meson remains massless in \schpt.
And in both cases the mass or mass-like terms give rise to interactions lacking enough derivatives to
suppress all IR divergences automatically.  Yet, one expects no IR divergences in the continuum
case as the strange quark mass turns on, and therefore
I expect \schpt\ to be likewise well-behaved.

However, a detailed proof that 
the chiral limit of full \schpt\ is ``safe'' may be rather delicate.  Although the 
lattice-dependent vertices give no nonderivative interactions
purely among the taste-$\xi_5$ 
mesons \cite{LEE-SHARPE,SCHPT1,Sharpe:2004is} (the analogous statement is also true
in the continuum example), 
potential problems may arise when external non-Goldstone mesons (taste different from $\xi_5$)
scatter to give massless $\xi_5$ mesons in internal loops. 

\begin{figure}[ht]
\resizebox{3.0in}{!}{\includegraphics{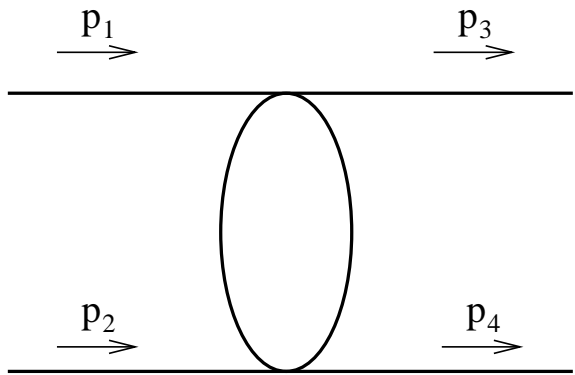}}
\caption{A ``t-channel'' scattering diagram for the scattering of two taste-singlet mesons,
with internal taste-$\xi_5$ mesons: $\pi^+_I(p_1)+\pi^-_I(p_2)\to\pi^+_I(p_3)+\pi^-_I(p_4)$.
\label{fig:scattering}}
\end{figure}

As a simple example, consider 
massless \schpt\ theory with two flavors,  
and take only the coefficient $C_4$ in the
taste-violating, $\cO(a^2)$, potential \cite{LEE-SHARPE,SCHPT1} to be nonzero. This is sufficient to
give mass to all the non-Goldstone mesons.  Now consider the contribution
shown in \figref{scattering} to the scattering of two taste-singlet, flavor charged, 
mesons through the exchange of taste-$\xi_5$ mesons.  If the two vertices come from the $C_4$
term, then there are no derivatives on the internal lines to suppress the logarithmic IR divergence.
On shell, the momentum transfer through the loop will, in the generic case,
provide a cutoff.  
However, with $C_4$ vertices, there is an IR divergence 
in this diagram in  the forward direction, $p_1=p_3$, $p_2=p_4$. The divergence is canceled
by the diagrams with one or both of the vertices replaced by kinetic energy vertices.
The terms in the kinetic energy vertices where the derivatives act on the external lines
provide the cancellation, possible because $p_1=p_3$ implies $p_1\cdot p_3 
= -a^2\Delta_I = -64a^2 C_4/f^2$.
Terms where the derivatives act on the internal lines are clearly not IR divergent; neither is the
corresponding ``s-channel'' diagram (on shell).
Though the result of this example is positive (no IR divergence on shell), it also shows some
of the issues that would need to be addressed in a complete proof that full \schpt\ is 
safe.  

A further subtlety is the following: When I suggest that full \schpt\ is ``safe,'' I really only mean
that certain standard quantities are safe, \ie they
have commuting chiral and continuum limits.  A sufficient
number of derivatives (with respect to quark mass) of any safe quantity will certainly produce
IR divergences and hence lack of commutativity of limits.  Which quantities are likely
to be safe?  
The above arguments suggest that quantities that are free of IR divergences in the chiral limit
of full continuum \chpt\ will probably be safe in \schpt. This means, for example, that pseudoscalar
masses and decay constants as computed in \cite{FPI04} will be safe.  
On the other hand, a quantity like the pion charge radius,\footnote{I thank
S.\ Sharpe for this example} 
which is defined
in terms of an external current with nonvanishing momentum transfer, is IR divergent
in the continuum chiral limit \cite{CHARGE-RADIUS}. The charge radius is therefore almost certain to
have the wrong chiral limit at fixed $a$ in \schpt;
it should indeed behave much the same way as the quenched $M_\pi^2/2\mu m_{\scriptscriptstyle V}$,
\eqs{CQmass}{SQmass}.

Furthermore, since in the continuum one can differentiate the squared meson mass once
with respect to the quark mass and still have a finite quantity in the chiral limit, I expect 
that the corresponding quantities in \schpt\ are safe.  For the squared Goldstone 
(taste $\xi_5$) mass this is fairly obvious from the
exact lattice $U_A(1)$; one can also  divide by (instead of differentiating with respect to)
the quark mass, as in the 
full \schpt\ version of \eq{SPQmass}.
It is less obvious that one can differentiate the squared masses of
non-Goldstone (taste other than $\xi_5$) mesons and still have a safe quantity. For example,
how do we know that a term like $a^2\chi\ln{\chi}$ cannot appear in the squared non-Goldstone
masses? Such terms are not divergent as $m\to0$, but their derivatives are divergent,
which would lead to noncommutativity. A one-loop calculation, similar to the scattering calculation described
above, shows  that such terms do in fact cancel for the taste-singlet
mass.  Whether this really persists beyond one loop remains to be seen.  

Note that dividing a squared non-Goldstone mass by quark mass, rather than differentiating, 
is certainly {\it unsafe}\/, as is already clear at tree-level, \eq{ChiVVI}. This 
shows why it is difficult to write down a simple rule that distinguishes safe from unsafe
quantities in all cases.  On the other hand, if we limit ourselves only to Goldstone mesons 
on the external lines, then it appears that \schpt\ quantities will be safe if
the corresponding quantities in full continuum \chpt\ are finite in the chiral limit.

\section{Errors of existing numerical computations}
\label{sec:errors}
D\"urr and Hoelbling \cite{Durr:2004ta} worry that the possible noncommutativity of limits with
staggered fermions would induce new, uncontrolled,  errors in the previously reported
results with MILC staggered configurations \cite{FPI04,strange-mass,MILC_SPECTRUM2,BIG_PRL}.
To the extent that \schpt\ describes the MILC simulations, we can argue that such errors
are not present --- even in the absence of a detailed proof that full 
\schpt\ is safe to all orders.
That is because the extraction of physical results requires extrapolation only to
the physical light quark masses, not to the chiral limit.  
The factor $\ln(\chi/\Lambda^2)$
is of order 3 or 4 at the physical point ($\chi\approx 140\,\MeV$).  
Even should the logarithm not be ``protected'' by 
powers of $\chi$ in some higher order of \schpt\ and 
represent a divergence in the chiral limit at fixed
$a$, it would still be  suppressed, numerically, 
by the additional powers of $a^2$ at that order.  By looking at the numerics
of the calculations in Refs.~\cite{FPI04,strange-mass,MILC_SPECTRUM2,BIG_PRL}, 
it is not hard to convince oneself that any new errors would most likely be
significantly smaller than the already-quoted systematic errors. 

As an example, consider the case of $f_\pi$.  Imagine that there is in fact an ``unsafe''
contribution at NNLO (next-to-next-to-leading order), \ie of order $a^4\ln(\chi/\Lambda^2)$.
To estimate its size, define the dimensionless taste-violating chiral expansion parameter 
$x_{a^2}$ by \cite{FPI04} 
\begin{equation}\label{eq:chia2def}
x_{a^2} \equiv \frac{a^2\overline{\Delta}}{8\pi^2 f_\pi^2} \ ,
\end{equation}
where $a^2\overline{\Delta}$ is a ``typical'' taste-violating term at $\cO(a^2)$. 
Taking for $a^2\overline{\Delta}$ the average pion splitting,  gives
$x_{a^2}\approx 0.09$ on the MILC coarse lattices and $x_{a^2}\approx 0.03$ on the fine lattices. 

Then I would expect the putative NNLO unsafe contribution to be generated by a term of size
\begin{equation}\label{eq:unsafe}
(x_{a^2})^2\; \frac{1}{16}\sum_B \ln{\chi_B/\Lambda^2}\ ,
\end{equation}
where the sum is over all $16$ pion tastes, labeled by $B$. The unsafe contribution comes
from the Goldstone pion in the sum.  Note that I assume that the Goldstone pion appears
only through such an average over tastes. That seems very likely, since the
diagrams that distinguish separate tastes are diagrams with disconnected internal
pion propagators, which
at NLO occur just for the singlet, axial, and vector tastes. At NNLO there could
be disconnected internal Goldstone lines, but it is known \cite{Sharpe:2004is} that 
such contributions come with an extra factor of $p^2$, making them safe.
Putting in 
$\ln{\chi/\Lambda^2}\approx4$ and
the values of $x_{a^2}$ gives an unsafe contribution of $\approx 0.2\%$ on the coarse
lattices and $\approx 0.02\%$ on the fine lattices. These are
clearly much smaller than other systematic errors (which are $\approx\!3\%$ for $f_\pi$
and $\approx\!1\%$ for $f_K/f_\pi$) , even without taking into account the
fact that some of the error would be removed by the continuum extrapolation 
(linear in $\alpha_S^2 a^2$ in Ref.~\cite{FPI04}).  

Of course, there would also be systematic errors coming from
contributions of the other taste pions in \eq{unsafe}.  However, these contributions
are safe, with commuting chiral and continuum limits. The associated errors
are therefore standard higher order \schpt\ errors, and as such are already included
in the error estimates of Ref.~\cite{FPI04}.

Note that my error estimate assumes that the ``worst case'' would be the existence of
a singularity in full \schpt\ that would interfere with the extrapolation to the
physical light quark masses.  This is reasonable despite the fact that intermediate
stages in the analysis in Refs.~\cite{FPI04,strange-mass,MILC_SPECTRUM2,BIG_PRL} often involve
fits to partially quenched lattice data, where we know that
noncommutativity of limits does occur.  The point is that the relevant 
NLO formulas
such as \eq{SPQmass} are known.  We fit lattice data to them in a range of valence and sea masses
that are far from the singularities. Since the light quark masses in 
the data are always more than a factor of three
larger than the physical masses,  the logarithms and the nonlinearities associated
with them are always significantly smaller than at the physical point.  It is therefore
not surprising that (as checked directly in the calculations) the NLO
corrections in the fits are under control and of the expected magnitude. If unexpected
new singularities were to occur in higher order, the resulting errors in the partially quenched fits
would be smaller than those induced in the subsequent full \schpt\ extrapolation, simply because
the logarithms are smaller.

Of course, if the staggered simulations had some additional noncommutativity of limits 
not captured by \schpt\ --- having to do, say, with the
fourth-root procedure --- then an uncontrolled
systematic error might in fact be present in the MILC
results.  However, recent advances in understanding the rooting procedure 
\cite{Maresca:2004me,Adams:2004mf,STAG-RG},
coupled with the good fit of MILC data \cite{FPI04} 
to \schpt\ predictions \cite{SCHPT1,SCHPT2},
make such a possibility seem increasingly unlikely.

\section{Additional Remarks}
\label{sec:remarks}
The lack of commutativity of the chiral and continuum limits is not exclusively a property
of staggered fermions.  For Wilson fermions, it appears at one loop in a power counting
for which $\cO(m)$ and $\cO(a^2)$ are treated as 
comparable \cite{Aoki:2003yv}. 
At this order, a term of the form $a^2\chi\ln{\chi}$ 
is present in the squared meson mass in the full QCD case.\footnote{I am 
grateful to O.\ B\"ar for pointing out to me this result in Ref.~\protect{\cite{Aoki:2003yv}}.} 
(The corresponding partially quenched case has not to my knowledge been studied.)
After differentiating with respect to (or dividing by) the quark mass, such a term
leads to an IR divergence
as $m\to0$ for fixed $a$; whereas it clearly vanishes if the $a\to0$ limit  is taken first.
The source of problem is the absence of an exact chiral symmetry in the massless
limit.  This means that massless Wilson pions can interact without derivative couplings
\cite{Kawamoto:1981hw}, leading to IR singularities and failure of commutativity.

Note that the Wilson case is significantly worse than the staggered case: In the latter, 
self-interactions of the massless Goldstone (taste $\xi_5$) pion are 
always proportional to at least two powers
of momentum.  At finite $a$, the non-Goldstone staggered pions are not required 
to have derivative interactions, but neither are they massless.

My \schpt\ examples and certainly the Wilson case suggest that lack of commutativity has little
or nothing to do with the rooting procedure.  Is this also true of the 2D Schwinger models
studied in Ref.~\cite{Durr:2004ta}?  The physics of the one-flavor 2D Schwinger model 
$\langle\bar\psi\psi\rangle$,
where the noncommutativity is found, is quite distinct from the multiflavor 4D chiral 
theories discussed above. With one flavor, the ``condensate'' comes about because
of the symmetry is violated by the anomaly, not because of spontaneous symmetry breaking.
In the continuum, the anomaly leads to exact zero modes of the Dirac operator, which
in turn saturate $\langle\bar\psi\psi\rangle$. 
Since staggered fermions have only near zero modes
in complex pairs, their effect cancels in the $m\to0$ limit, and  $\langle\bar\psi\psi\rangle$ 
vanishes for fixed $a$. If the continuum limit is taken first, however, the complex 
pair of near zero modes
becomes a pair of degenerate exact zero modes, and the rooting prescription works as desired, producing
a single exact mode.
(See Refs.~\cite{Durr:2004ta,Durr:2003xs} for details.)  Since the physics here is
inextricably tied to having only one flavor, however, it is not possible to separate cleanly the
issues of the lack of commutativity and the rooting procedure within this model.  

In the two-flavor 2D Schwinger model, on the other hand, the rooting procedure is not required,
yet I suspect that it would be possible to find noncommutativity in the masses of the
``quasi-Goldstone bosons,'' of a similar nature
to that seen in quenched or partially quenched 4D \schpt\ (\secref{examples}). That is because
integration over a single (boson) pole in 2D has the same IR behavior as integration over
a double pole in 4D.  The problem here is that there is no true condensate in 2D, so there
is no simple chiral theory: $\langle\bar\psi\psi\rangle$ 
and the quasi-Goldstone boson squared masses
vanish as fractional powers of $m$ as $m\to0$ \cite{COLEMAN}. 
It is therefore a nontrivial problem to discover an analogue of \schpt\ that would allow one to
calculate the discretization effects caused by staggered fermions. Thus I am not able
at this point to make a specific proposal for where to look for noncommutativity in this
model.

A simple alternative to the two-flavor theory in this context can be
found in the quenched 2D Schwinger model, which does not require the rooting procedure
either (trivially).  Reference~\cite{Durr:2003xs} studies the quenched
theory with staggered fermions, but since $\langle\bar\psi\psi\rangle$ blows up as $1/m$
in the continuum, it is hard to see any noncommutativity cleanly. Recently,
D\"urr and Hoelbling have looked instead 
at $m\langle\bar\psi\psi\rangle/g^2$ ($g$ is the coupling)
for staggered fermions \cite{DURR-PRIVATE}.  They find clear evidence that the chiral and continuum
limits do not commute for this quantity.  In addition, the behavior is
is very similar, qualitatively,  to that observed for 
$\langle\bar\psi\psi\rangle$ in the one-flavor Schwinger model.  At the least, this shows
that noncommutativity of limits is not inextricably tied to the rooting procedure in these
2D models.

Finally, one may wonder whether there are order-of-limits problems
in 4D that are directly analogous to those in the one-flavor 2D Schwinger model, and not
of the kind treated above in \schpt.   For example, in
QCD with a single flavor, the value of $\langle\bar\psi\psi\rangle$  
in the chiral limit is probably crucially dependent on having exact zero 
modes \cite{Durr:2003xs}, just as it is in 2D, leading to noncommuting chiral and continuum limits.   
In normal, multiflavor QCD, however, I know of no standard physical quantities
that have this kind of sensitivity to the existence of exact zero modes.   For example,
the topological susceptibility seems to be well-behaved, both in
\schpt\ \cite{Billeter:2004wx} and in simulations \cite{MILC-TOPO}.  This is not
to say that the staggered discretization errors are negligible --- indeed they
are large.  But there is no commutativity problem; the continuum suppression
of the susceptibility seems to be present whether one takes the chiral or continuum limit first. 
Similarly, I do not expect any problem in the QCD $\eta'$ mass. 
Because of the anomaly, the
$\eta'$ mass is generated through disconnected meson diagrams \cite{Witten:1979vv}. 
In the continuum limit, staggered fermions have the correct anomaly (for infinitesimal chiral
rotations) \cite{Sharatchandra:1981si}, and this
should be adjusted correctly by the fourth-root procedure, since the anomaly is essentially
perturbative. At finite lattice spacing, 
I expect that \schpt\ will correctly capture the modifications
in the $\eta'$ physics due to staggered fermions, 
at least at vacuum angle
$\theta=0$,\footnote{Away from $\theta=0$, the issue is not simply one
of commutativity of limits, but whether staggered fermions can get the correct physics at all.
This is related to the fact that the 
staggered anomalous symmetry involves a lattice translation
and cannot be exponentiated to trade 
a phase  in the quark mass matrix for $\theta$.
For example, the known 
CP-violating phase \protect{\cite{NEG-MASSES-CP}}
cannot be reproduced by a simulation with an
odd number of negative mass staggered quarks \protect{\cite{Creutz:2003xu}}.}
just as ordinary \chpt\ correctly captures $\eta'$ physics in the continuum \cite{GASSER_LEUTWYLER}.

\section*{ACKNOWLEDGMENTS}
I am very grateful to Stephan D\"urr and Christian Hoelbling for extensive discussions
and for communicating their results on the quenched Schwinger model before publication.
I also thank Oliver B\"ar, Maarten Golterman, Jim Hetrick, Gautam Rupak, and Steve Sharpe for many
helpful comments and suggestions.  Finally, I am grateful to Peter Weisz for
pointing out the misleading description of ``standard quantities'' in version 2 of this paper. This work is
partially supported by the US Department of Energy under grant
DE-FG02-91ER40628.

\end{document}